\documentstyle[psfig]{l-aa}

\begin{document}

 \title{
A strange star model for GRO J1744-28
}

 \author{
R.X. Xu\inst{1,3},
G.J. Qiao\inst{2,1,3}
        }

 \institute{ Department of Geophysics, Peking University, Beijing 100871, China
   \and
         CCAST (World Laboratory) P.O. Box 8730, Beijing 100080, China
   \and
         Chinese Academy of Sciences-Peking University joint Beijing Astrophysical Center, Beijing 100871, China
           }
 \thesaurus{08.16.6}
 \date{Received \hspace{2cm} ; accepted \hspace{2cm}}
 \maketitle
 \begin{abstract}

A model is suggested for the bursting x-ray pulsar, GRO J1744-28, where the compact object of the pulsar is proposed to be a strange matter star with crusts. It is suggested that two envelope crusts shield the two polar caps of an accreting strange stars owing to strong polar-cap magnetic fields. GRO J1744-28 might not have a unified crust, but two polar crusts. Bursts are the result of phase transition of polar crusts if one accreting crust is heavier than that the Coulomb force can support. It is shown that there should be phase-lags when bursting crust is expanded about one hundred meters. Many calculated properties of this model, such as the phase-lags, the bursting luminosity and the characteristic photon energy, are in agreement with observations.


 \end{abstract}
%

\section{Introduction}

GRO J1744-28 is the first known X-ray source both bursting and pulsating (see Rappaport \& Joss 1997, Aptekar et al. 1998, and references therein). The source is in a binary system, the X-ray pulsation period is 467 ms, the orbital period is 11.8 days, the mass function of this system is $1.31\times 10^{-4} {\rm M}_\odot$, the time interval between bursts is about 30 minutes, and the duration of the bursts is about 5 seconds. There is not any distinct difference between the spectra of the bursts and persistent emission, and the typical photon energy is about 14 keV. When bursting, there are post-burst dips in the persistent flux and phase-lags. The phase of the pulse profile during bursts lags the phase prior to the bursts by $70^o$-$50^o$.

Any theoretical model should make clear that, such as, how matter accretes onto the surface of GRO J1744-28, how the bursts produce, why there are phase-lags and post-burst dips, etc. Usually, people believe that the source could be well explained by a complex accreting process onto neutron star. One of this kind of model believe that the bursts are the result of accretion instability (Type II X-ray bursts, Lewin et al. 1996), which has been studied in the Rapid burster (MXB 1730-335). On the oher hand, Bildsten \& Brown (1997) show that the bursts observed during the peak of the outburst are most likely not the thermonuclear origin, i.e., GRO J1744-28 could not be Type I X-ray burster.

However, Cheng et al. (1998) have shown the possibility that the burst energy is the result of phase transition of accreted matter from normal hadron to three-flavor strange quark matter. Hence, they concluded that GRO J1744-28 might be an accreting strange star. In their discussion, they address that GRO J1744-28 is a strange star with a normal matter crust. When the accreted mass exceeds some critical mass, the crust breaks, resulting in the phase transition and the energy release.

Xu \& Qiao (1998) argued that the normal matter crust is not necessary for the strange star model of pulsars. If a strange star (Alcock et al. 1986) forms soon after a supernova, a magnetosphere mainly composed by $e^\pm$ would soon be established, and the strange star could act as a radio pulsar. There are some advantages if radio pulsars are bare strange stars rather than neutron stars or strange stars with crust. While strange stars are in binaries, there could be two envelope crusts shielding the two polar caps of the accreting strange stars.

Following this idea, we propose a model for GRO J1744-28, which is similar to that by Cheng et al.(1998). We argue that GRO J1744-28 might not own a unified crust, but two polar crusts. Bursts are the result of phase transition of polar crusts if one accreting crust is heavier than that the Coulomb force can support. This phase transition process could be unstable, similar to the case discussed by Horvath \& Benvenuto (1988). It is shown that there should be phase-lags when bursting crust is expanded about one hundred meters, a scale length in order of the crust thickness.

About the phase-lag, Miller (1996) has suggested that when accretion rate rises, the flattened accretion flow shift to a different set of field lines. Therefore, the time of pulse maximum would change accordingly, and a phase-lag results. Our model is different from this scenario.

\section{General Description of the Crust over Strange Quark Matter}

As $s$ quark is slightly heavier than $u$ and $d$ quarks, there are $u$, $d$, $s$ quarks and electrons for the chemical and electrical equilibrium of strange quark matter. Since the quark matter are bound up through strong interaction, and the electrons are held to the quark matter electrically, the electrons' distribution would extend beyond the quark matter surface several hundred fm, and there is a strong electric field near the quark surface. Using the Thomas-Fermi model, one can obtain the electric field $E$ as a function of the height above the quark surface $z$ (Xu \& Qiao, 1998, equ. A5-A6),
$$
\begin{array}{lll}
E = -{dV \over dz} & = & \sqrt{2\alpha\over 3\pi} \cdot {9 V_q^2 \over
    (\sqrt{6\alpha\over \pi} V_q \cdot z + 4)^2}\\
& \sim & {1.42 \times 10^{21} \over (1.2 z_{11} + 4)^2} \;\; {\rm V\;m^{-1},}
\end{array}	\eqno(1)
$$
where, the direction of the electric field is outward, and the potential energy of quark matter $V_q$ has been chosen to be 20 MeV, $z_{11} = z/(10^{-11}$ cm).

As matter accreting on to the polar caps, the strange quark matter should repulse atomic nuclei, and two crusts are formed in the two polar caps. Assuming the density of the crust is $\rho_{\rm crust}$, the distance $l$ between nuclei is
$$
\begin{array}{lll}
l & = & ({A\times m_{\rm p} \over \rho_{\rm crust}})^{1/3}\\
& \sim & 1.0 \times 10^{-11} A_{56}^{1/3} \rho_{11}^{-1/3} \;\; {\rm cm,}
\end{array}
$$
where $m_{\rm p}$ is proton mass, $A$ is the mass number of nucleus, $A_{56}=A/56$, $\rho_{11}=\rho_{\rm crust}/10^{11}$g cm$^{-3}$. If $\rho_{\rm crust}$ approximates neutron drip density, $\rho_{11} \sim 1$ (Glendenning et al. 1995), then $l$ is order of the electric gap height(Alcock et al. 1986).

By simulation, Huang \& Lu (1997) argued that the electric gap between crust and strange quark matter is too small to support the crust before the density at the bottom rises to the neutron drip density. Because of the repulsion between nuclei and strange quark matter, there should be a charge separation near the bottom of crust. Hence there would be an effective positively charged column density $\sigma_{\rm e}$, which could be estimated as
$$
\begin{array}{lll}
\sigma_{\rm e} & = & \eta\cdot {Z e \over l^2}\\
& \sim & 2.7 \times 10^{23} \eta \rho_{11}^{2/3} A_{56}^{-2/3} Z_{26} e,
\end{array}
$$
where $\eta \sim 1$ is a coefficient denoting the effective (positive) chargement, $Z$ is charge number of atomic nucleus, $Z_{26} = Z/26$, $e$ is the charge of electron. Hence, by equilibrium between electrical and gravitational forces, the column density $\sigma_{\rm m}$ of crust can be written as
$$
\begin{array}{lll}
\sigma_{\rm m} & = & {\sigma_{\rm e} E R^2 \over G M_*}\\
& \sim & {4.6 \times 10^{16} \over (1.2 z_{11} + 4)^2} \eta \rho_{11}^{2/3} A_{56}^{-2/3} Z_{26} M_1^{-1} R_6^2 \;\; {\rm g\cdot cm^{-2}}\\
& \sim & {2.3 \times 10^{-17} \over (1.2 z_{11} + 4)^2} \eta \rho_{11}^{2/3} A_{56}^{-2/3} Z_{26} M_1^{-1} R_6^2 \;\; {\rm M_{\odot}\cdot cm^{-2}},
\end{array}	\eqno(2)
$$
where $M_*$ and $R$ are the mass and radius of strange quark matter, respectively, $M_1 = M_*/(1M_{\odot})$, $G$ is the gravitational constants. The height $H_{\rm crust}$ of the crust is
$$
\begin{array}{lll}
H_{\rm crust} & \sim & {\sigma_{\rm m} \over \rho}\\
& \sim & 1.1 \times 10^4 \eta \rho_{11}^{-1/3} A_{56}^{-2/3} Z_{26} M_1^{-1} R_6^2 \;\; {\rm cm \;\;(for\;} z_{11}=2).
\end{array}
$$
For a spherical crust, the total mass $M_{\rm crust}$ should be
$$
\begin{array}{lll}
M_{\rm crust} & = & 4\pi R^2\sigma_{\rm m}\\
& \sim & {29 \times 10^{-5} \over (1.2 z_{11} + 4)^2} \eta \rho_{11}^{2/3} A_{56}^{-2/3} Z_{26} M_1^{-1} R_6^4 \;\; {\rm M_{\odot}},
\end{array}
$$
which is a typical value of simulated results (e.g. Huang \& Lu 1997). For $z_{11}=0$ (the length scale of strong interaction is $\sim z_{11}=10^{-2}$), we get the maximum values of column density $\sigma_{\rm max}$ and of spherical total mass $M_{\rm max}$,
$$
\begin{array}{lll}
\sigma_{\rm max} & = & 1.4 \times 10^{-18} \eta \rho_{11}^{2/3} A_{56}^{-2/3} Z_{26} M_1^{-1} R_6^2 \;\; {\rm M_{\odot}\cdot cm^{-2}},\\
M_{\rm max} & = & 1.8 \times 10^{-5} \eta \rho_{11}^{2/3} A_{56}^{-2/3} Z_{26} M_1^{-1} R_6^4 \;\; {\rm M_{\odot}}.
\end{array}	\eqno(3)
$$

If the mass of the crust is known, the distance between the crust and strange quark matter should be (from equ.(2))
$$
z_{11} \sim  4.0 \eta^{1/2} \sigma_{18}^{-1/2} \rho_{11}^{1/3} A_{56}^{-1/3} Z_{26}^{1/2} M_1^{-1/2} R_6 - 3.3,
$$
where $\sigma_{18} = \sigma_{\rm m}/(10^{-18} {\rm M}_{\odot}{\rm cm^{-2}})$. It is shown from above equation that the crust becomes heavier as matter accretes, the thickness of the electric gap becomes smaller.

\section{Bursting Process above the Polar Cap}

As matter accreting to the polar cap crusts, the column density of crusts could be more and more higher, until a catastrophe of crust takes place when $\sigma_{\rm m} > \sigma_{\rm max}$, because the strong interaction length is of order 1 fm. This crush of crust could trigger a burst process in GRO J1744-18.

When a crush happens, the strange quark core should swallow part of normal matter at the bottom of the crust. This process could not be unstable (possibly in a detonation mode), which might be similar to that discussed by Horvath \& Benvenuto (1988).

Before going to some details of the model for GRO J1744-28, we would like to briefly review the three types of particular bursts. There are two types of X-ray bursts, which we call them as type I and type II bursts(Lewin et al. 1993). The ratios of steady emission to time averaged burst emission in type I bursts, $\alpha_{\rm I}$, and in type II bursts, $\alpha_{\rm II}$, are
$$
\begin{array}{lllll}
\alpha_{\rm I} & \sim & {10^{20} {\rm ergs\cdot s^{-1}} \over 10^{18} {\rm ergs\cdot s^{-1}}} & \sim & 100,\\
\alpha_{\rm II} & \sim & {1.3 \times 10^{37} {\rm ergs\cdot s^{-1}} \over 1.3 \times 10^{37} {\rm ergs\cdot s^{-1}}} & \sim & 1,
\end{array}
$$
respectively, while, the Eddington luminosity $L_{\rm Edd} \sim 10^{38} {\rm ergs\cdot s^{-1}}$. For the bursts in GRO J1744-18, in the early 1995 Dec., when the burst intervals were $\sim 3$ minutes, the $3\sigma$ upper limit of the ratio of integrated energy in the persistent flux to that in bursts was $\sim 4$ (Lewin et al. 1996), i.e.,
$$
\alpha_{1744} \leq 4.
$$

There are three main energy sources in the X-ray emission by unit nucleon. First, the gravitational accretion energy, $\varepsilon_{\rm G}$. Second, the thermonuclear reaction energy from pure hydrogen into helium, $\varepsilon_{\rm H}$, and from helium into iron-peak elements, $\varepsilon_{\rm He}$. Third, the energy liberated in the phase transition from normal hadron into strange quark matter, $\varepsilon_{\rm s}$. They are
$$
\begin{array}{lll}
\varepsilon_{\rm G} & \sim & 140 {\rm MeV/nucleon},\\
\varepsilon_{\rm H} & \sim & 6.7 {\rm MeV/nucleon},\\
\varepsilon_{\rm He} & \sim & 1.7 {\rm MeV/nucleon},\\
\varepsilon_{\rm s} & \sim & 30 {\rm MeV/nucleon}.
\end{array}
$$

For type I X-ray bursts caused by nuclear flash, the observed ratios $\alpha_{\rm I}$ are from $\sim {\varepsilon_{\rm G} \over \varepsilon_{\rm H} + \varepsilon_{\rm He}}$ to $\sim {\varepsilon_{\rm G} \over \varepsilon_{\rm He}}$. For type II X-ray bursts, which are the results of accretion instability, the observed ratios $\alpha_{\rm II} \sim {\varepsilon_{\rm G} \over \varepsilon_{\rm G}} \sim 1$ if the time averaged accretion rates during bursts and between bursts are similar.

While, for the busting X-ray pulsar, GRO J1744-28, there should be a phase transition from hadron to three-flavor quark matter when the crust crashes. Nevertheless, the energy release from the phase transition could trigger the thermonuclear reactions in the crust. Hence, the ratio of steady emission to time averaged burst emission could be ${\varepsilon_{\rm G} \over \varepsilon_{\rm s} + \varepsilon_{\rm H} + \varepsilon_{\rm He}} \sim 3.6$, which is comparable to the observed $\alpha_{1744}$.

The polar cap area of the crust, $A_{\rm p}$, is
$$
A_{\rm p} = {\pi R^3 \over r_{\rm e}},
$$
where, $r_{\rm e} = {\rm min}[r_{\rm c}, r_{\rm m}]$, and $r_{\rm c}$ is the radius of light cylinder, $r_{\rm m}$ is the radius of magnetosphere (Rappaport \& Joss, 1997),
$$
\begin{array}{lll}
r_{\rm c} & = & {cP\over 2\pi} = 2.2\times 10^9 \;\; {\rm cm},\\
r_{\rm m} & = & 2.8\times 10^7 B_{11}^{4/7} R_6^{12/7} M_1^{-1/7} \stackrel{.}M_8^{-2/7} \;\; {\rm cm}.
\end{array}
$$
Where $B$ is the magnetic field, $B_{11}=B/(10^{11}{\rm gauss})$, $\stackrel{.}M$ is the accretion rate, $\stackrel{.}M_8 = {\stackrel{.}M \over 10^{-8} M_{\odot}\cdot {\rm yr^{-1}}}$, $P = 0.467$s is the period of rotation, $c$ is the velocity of light. Hence (let $r_{\rm e}=r_{\rm m}$),
$$
A_{\rm p} = 1.1\times 10^{11} B_{11}^{-4/7} R_6^{9/7} M_1^{1/7} \stackrel{.}M_8^{2/7} \;\; {\rm cm^2}.	\eqno(4)
$$

The typical interval between bursts is about 30 minutes (Aptekar et al. 1998). We assume the bursting and accreting processes are in quasi-equilibrium, and the total energy release during a burst is
$$
\begin{array}{lll}
E_{\rm tot} & = & {\stackrel{.}M\times 30 {\rm minutes} \over m_{\rm p}} \times 30 {\rm MeV}\\
& \sim & 2.1\times 10^{46} \stackrel{.}M_8\;\; {\rm MeV}\\
& \sim & 3.3\times 10^{40} \stackrel{.}M_8\;\; {\rm ergs}.
\end{array}	\eqno(5)
$$
The durative time during bursts $\delta T \sim 5$s (Aptekar et al. 1998), then the busting luminosity $L_{\rm burst} \sim {E_{\rm tot}\over \delta T} \sim 6.6\times 10^{39} \stackrel{.}M_8 \;{\rm ergs \cdot s^{-1}} \sim L_{\rm Edd}$. Therefore, the crust should be expanded very much, which could be the reason for the observed phase-lag (see section 4). The effective temperature $T_{\rm eff}$ is
$$
\begin{array}{lll}
T_{\rm eff} & = & ({L_{\rm burst} \over \sigma A_{\rm p}})^{1/4}\\
& \sim & 1.8\times 10^8 B_{11}^{1/7} R_6^{-9/28} M_1^{-1/28} \stackrel{.}M_8^{5/28} \;\; {\rm K}\\
& \sim & 16 B_{11}^{1/7} R_6^{-9/28} M_1^{-1/28} \stackrel{.}M_8^{5/28} \;\; {\rm keV}.
\end{array}	\eqno(6)
$$
where $\sigma = 5.67\times 10^{-5} {\rm ergs \cdot s^{-1}cm^{-2}K^{-4}}$ is the Stefan-Boltzmann's constant. While, in observation for GRO J1744-28, X-rays from 10 to 100 keV have been detected.

Also, the column density $\sigma_{\rm crush}$ of phase transition matter in the crush of crust should be
$$
\begin{array}{lll}
\sigma_{\rm crush} & = & {\stackrel{.}M\times 30 {\rm minutes} \over A_{\rm p}}\\
& \sim & 5.2\times 10^{-24} B_{11}^{4/7} R_6^{-9/7} M_1^{-1/7} \stackrel{.}M_8^{5/7}\;\; {\rm M_{\odot}\cdot cm^{-2}}\\
& \sim & 1.0\times 10^{10} B_{11}^{4/7} R_6^{-9/7} M_1^{-1/7} \stackrel{.}M_8^{5/7}\;\; {\rm g\cdot cm^{-2}},
\end{array}
$$
and the extent scale $h_{\rm crush}$ of phase transition matter can be estimated as
$$
\begin{array}{lll}
h_{\rm crush} & = & {\sigma_{\rm crush} \over \rho}\\
& \sim & 0.1 B_{11}^{4/7} R_6^{-9/7} M_1^{-1/7} \stackrel{.}M_8^{5/7}\rho_{11}^{-1}\;\; {\rm cm} \ll H_{\rm crust},
\end{array}
$$
which means only a small faction of matter in the crust has been transformed from hadron into three-flavor quark matter.

\subsection{The Bursting Process is Unstable}

In equ.(3), it is said that $\sigma_{\rm max} = \sigma_{\rm m} (z_{11}=0)$. In fact, as the temperature of the crust is not zero, ions can penetrate a small potential barrier with height of $\sim kT$, where $k$ is the Boltzmann constant, $T$ is the temperature of matter at the base of the crust. From equ.(A6)
$$
V={3V_q\over \sqrt{6\alpha\over\pi}V_qz+4},\;\;({\rm for} z > 0)
$$
of Xu \& Qiao (1998), we get the differential relation of $V$ and $z$ as
$$
\begin{array}{lll}
\delta V & = & -{3V_q\over (\sqrt{6\alpha\over\pi}V_qz+4)^2} \times \sqrt{6\alpha\over\pi}V_q \delta z_{11}\\
& \sim & -4.5 \delta z_{11},
\end{array}
$$
for $V_{\rm q}=20$MeV, $z_{11}=0$. If $Z \delta V\sim kT\sim 10^{-2}$MeV (the charge number $Z$ of ion is chosen to be 26), then $\delta z_{11}\sim 10^{-4}\ll 1$. From equ. (2), the differential relation of $\sigma_{\rm m}$ and $z$ is
$$
\delta \sigma_{\rm m} \sim -8.6\times 10^{-19} \delta z_{11} \eta \rho_{11}^{2/3} A_{56}^{-2/3} Z_{26} M_1^{-1} R_6^2,
$$
for $V_{\rm q}=20$MeV, $z_{11}=0$. Hence, if $kT \sim 10^{-2}$MeV, $\delta \sigma_{\rm m} \sim 10^{-23} \ll \sigma_{\rm m}$, which means equ. (3) is a good approximation for the maximum value of column density.

When the accreting crust is heavier than that the Coulomb force can support, the bottom part of the crust should be pushed through the Coulomb barrier. The released energy from phase transition would heat up the bottom crust, and according to the differential relations of $V-z$ and $\sigma_{\rm m}-z$, more ions should penetrate the Coulomb barrier. In a word, a positive feedback process happens. If the temperature in the feedback process (and {\it before} the crust have taken a global expansion) is risen to be $T+\delta T$, while $\delta T$ is assumed to be order of $10^{-3}$ MeV, the total mass been pushed through should be order of $10^{-24}$ ({\it note}: $\sigma_{\rm crush} \sim 10^{-24}$) $M_\odot \cdot$ cm$^{-2}$.

\section{The Phase-lag}

As discussed below, the bursting can cause matter being expanded large enough to drag matter to latter phase, and the observed phase-lag can be explained.

The total mass of one polar cap crust $M_{\rm polar}$ is
$$
\begin{array}{lll}
M_{\rm polar} & = & \sigma_{\rm m} A_{\rm p}\\
& \sim & {5.1 \over (1.2 z_{11} + 4)^2}\times 10^{27} \eta \rho_{11}^{2/3} A_{56}^{-2/3} Z_{26} M_1^{-6/7} R_6^{23/7} B_{11}^{-4/7} \stackrel{.}M_8^{2/7} \;\; {\rm g}.
\end{array}
$$
Let $\alpha$ be the inclination angle (Fig.1), the total rotational energy of the polar crust is
$$
\begin{array}{lll}
J_{\rm polar} & = & {1\over 2} M_{\rm polar} \Omega_0^2 R^2\sin^2\alpha\\
& \sim & {4.6\over (1.2 z_{11} + 4)^2}\times 10^{41} \eta \rho_{11}^{2/3} A_{56}^{-2/3} Z_{26} M_1^{-6/7} R_6^{37/7} B_{11}^{-4/7} \stackrel{.}M_8^{2/7}\sin^2\alpha \;\; {\rm ergs},
\end{array}
$$
where $\Omega_0 = {2\pi \over P}$ is the angular velocity of the star. When bursting, the matter of crust could be expanded to a maximum height $r$ (Fig.1). Assuming this expanded crust has a uniform angular velocity $\Omega$, from rotational energy conservation,

\begin{figure}[h]
\centerline{\psfig{figure=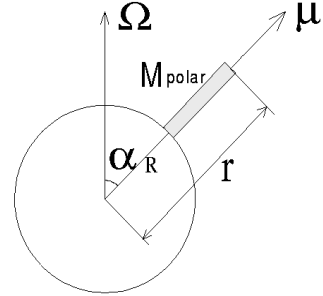,angle=0,height=4cm,width=4.2cm}}
\caption{
A demonstration of bursting process of accretion polar-crust. ${\bf \Omega}$ is the rotational axis, ${\bf \mu}$ is the magnetic axis, $\alpha$ is the inclination angle, and $R$ is the radius of strange star. When bursting, the crust would expanded to a maximum height $r$.
}
\end{figure}

$$
J_{\rm polar} = {1\over 6} M_{\rm polar} \Omega^2 (r^2+Rr+R^2) \sin^2\alpha,
$$
we get
$$
\Omega = \Omega_0 \sqrt{{3R^2\over r^2+Rr+R^2}}.
$$
If there is a expansion in the crust, $r\neq R$, then $\Omega_0\neq \Omega < \Omega_0$. Therefore, in the corotation frame, the crust matter would be drag to a later phase, and the total drag kinetic energy $E_{\rm k}$ is
$$
\begin{array}{lll}
E_{\rm k} & = & \int_R^r{1\over 2} ({{M_{\rm polar}\over r-R}}dx) [(\Omega_0-\Omega)x \sin^2\alpha]^2\\
& = & {1\over 6} M_{\rm polar} \Omega_0^2\sin^2\alpha(r^2+Rr+R^2)(1-\sqrt{3R^2\over r^2+Rr+R^2})^2.
\end{array}
$$
While, in the corotation frame, the magnetic force for matter to get back is $F = 2\cdot {B^2\over 4\pi} \cdot A_{\rm p}$ since the Maxswell tense force is ${B^2\over 4\pi}$ (Alfven \& Falthammar 1963). So, the shifted distance can be estimated as
$$
\begin{array}{lll}
S & \sim & {E_{\rm k} \over F}\\
& = & {\pi \over 3B^2A_{\rm p}} M_{\rm polar} \Omega_0^2\sin^2\alpha(r^2+Rr+R^2)\\
& & \;\; (1-\sqrt{3R^2\over r^2+Rr+R^2})^2\\
& \sim & {8.8\times 10^8 \over (1.2 z_{11} + 4)^2} f(r) \sin^2\alpha \eta \rho_{11}^{2/3} A_{56}^{-2/3} Z_{26} M_1^{-1} R_6^4 B_{11}^{-2} \;\; {\rm cm},
\end{array} \eqno(7)
$$
here the function $f(r)$ is
$$
f(r) = [1+{r\over R}+({r\over R})^2] \cdot (1-\sqrt{3R^2\over r^2+Rr+R^2})^2. \;\; \eqno(8)
$$
The degree of phase-lag $\theta$ is (for $z_{11}=0$),
$$
\begin{array}{lll}
\theta & = & {180^o \over \pi}\times {S\over R\sin\alpha}\\
& \sim & 3.2^o \times 10^3 \sin\alpha f(r) \eta \rho_{11}^{2/3} A_{56}^{-2/3} Z_{26} M_1^{-1} R_6^3 B_{11}^{-2}.
\end{array} \eqno(9)
$$
The numerical calculation of $f(r)$ show that $\theta$ could be consistent with observed phase-lags if $r \sim 1.01 R$ or $r-R=100$ meters.

\section{Conclusion and Discussion}

A model for GRO J1744-28 is suggested in this paper, with which many observational properties, such as the phase-lags, the burst energy and the characteristic photon energy, could be explained well and simply. There might be two polar cap crusts of GRO J1744-28. The intermittent phase transition of polar crusts could result in the burst process when an accreting crust is heavier than that the Coulomb force can support.

Bildsten \& Brown (1997) argue that the bursts of GRO J1744-28 during the peak of the outburst are most likely not the thermonuclear origin. Nevertheless, if the energy release by phase transition is considered, the thermonuclear reactions could have contribution to the bursting process in GRO J1744-28. Taam \& Picklum (1978, 1979) recognized that if hydrogen burns at sufficiently high density ($\sim 10^6$ g cm$^{-3}$) in the envelop of an accreting neutron star, helium must eventually burn within the same mass layer. This mixed burn of hydrogen and helium could occur in the accreting crust of GRO J1744-28, which might be triggered by the energy release of phase transition. Eventually, a burst appears, with a global expansion of $\sim 100$ meters in the crust.

Rappaport \& Joss (1997) think that X-ray pulsars, GRO J1744-28, the rapid Burster, and type I X-ray bursters are in a series when the crucial parameter B (magnetic field) decreases. However, the magnetic field of the recently discovered bursting X-ray millisecond pulsars, SAX J1808.4-3658,  is only $\sim 10^8$ Gauss (Wijnands \& Klis 1998). The spectra of GRO J1744-28 and SAX J1808.4-3658 are hard. The characteristic photon energy of GRO J1744-28 is comparable to most other X-ray pulsars but higher than other X-ray bursters. On the other hand, SAX J1808.4-3658 has a remarkably stable Crab-like power law spectrum (Gilfanov et al. 1998). Where and why is SAX J1808.4-3658 in the series of Rappaport \& Joss (1997)? We have to deal with the radiative mechanism of X-ray emission in GRO J1744-28 and SAX J1808.4-3658. The general standard pulsar high-energy emission mechanism could not work well. Nevertheless, a revised version of the standard mechanism considering the electromagnetic effect of accretion disk could be a good candidate-model for the radiative process of bursting pulsars. Detailed research in this scenario should be necessary.

\acknowledgements{We sincerely thank Prof. K.S. Cheng, Prof. T.Lu and Dr. Z.G. Dai for very helpful discussions. This work is supported by the National Nature Sciences Foundation of China, by the Climbing project of China, and by the Youth Foundation of Peking University.
}


\begin{thebibliography}{}

\bibitem[]{}
  Alfven H. \& Falthammar C.-G., 1963, {\it Cosmical Electrodynamics}, Oxford University Press
\bibitem[]{}
  Aptekar R.L., et al., 1998, ApJ 493, 404
\bibitem[]{}
  Alcock C., Farhi E., Olinto A., 1986, ApJ 310, 261
\bibitem[]{}
  Bildsten L., Brown E.F., 1997, ApJ 477, 897
\bibitem[]{}
  Cheng K.S., Dai Z.G., Wei D.M., Lu T., 1998, Science 280, 407
\bibitem[]{}
  Gilfanov M., Revnivtsev M., Sunyaev R., Churazov E., 1998, A\&A submitted
\bibitem[]{}
  Glendenning N.K., Kettner Ch., Weber F., 1995, ApJ 450, 253
\bibitem[]{}
  Horvath J.E., \& Benvenuto O.G., 1988, Phys. Lett. B213, 516
\bibitem[]{}
  Huang Y.F., \& Lu T., 1997, A\&A 325, 189
\bibitem[]{}
  Lewin W.H.G., et al., 1996, ApJ 462, L39
\bibitem[]{}
  Lewin W.H.G., van Paradijs J., Taam R.E., 1993, {\it Space Sci. Rev.} 62, 223
\bibitem[]{}
  Miller G.S., 1996, ApJ 486, L29
\bibitem[]{}
  Rappaport S., \& Joss P.C., 1997, ApJ 486, 435
\bibitem[]{}
  Taam R.E., \& Picklum R.R., 1978, ApJ 224, 210
\bibitem[]{}
  Taam R.E., \& Picklum R.R., 1979, ApJ 233, 327
\bibitem[]{}
  Wijnands R., \& van der Klis M., 1998, Nature submitted (astro-ph/9804216)
\bibitem[]{}
  Xu R.X., \& Qiao G.J., 1998, ApJ submitted (astro-ph/9804278)


 \end{thebibliography}
\end{document}